%% file: arxiv.tex
\documentclass{article} %
\usepackage{arxiv_style,times}

\input{math_commands.tex}

\usepackage{hyperref}
\usepackage{url}
\usepackage{xcolor}         %
\usepackage{siunitx}

\usepackage{graphicx} 
\usepackage{amsmath}
\usepackage{algorithm}
\usepackage{algorithmic}
\usepackage{cleveref}
\usepackage{enumitem}
\usepackage{subcaption}
\usepackage{amsthm}
\usepackage{xspace}
\usepackage{booktabs}
\usepackage{multirow}
\usepackage{float}

\usepackage{tcolorbox}
\tcbuselibrary{skins,breakable}

\newtcolorbox{fancybox}[2][]{%
  colback=gray!05,
  colframe=gray!10,
  coltitle=black,
  boxrule=0pt,
  title={#2},
  fonttitle=\bfseries,
  boxsep=2pt,
  left=2pt,
  right=2pt,
  top=2pt,
  bottom=2pt,
  #1 %
}

\newcommand{\methodname}{\textsc{SimKey}\xspace}

\newtheorem{theorem}{Theorem}[section]
\newtheorem{lemma}[theorem]{Lemma}

\title{SimKey: A Semantically Aware Key Module for Watermarking Language Models}

\author{
Shingo Kodama\textsuperscript{1*},
Haya Diwan\textsuperscript{2*},
Lucas Rosenblatt\textsuperscript{2},
R.~Teal Witter\textsuperscript{3},
Niv Cohen\textsuperscript{2}\\
\textsuperscript{1}Middlebury College \quad
\textsuperscript{2}New York University \quad
\textsuperscript{3}Claremont McKenna College
}

\begingroup

\footnotetext[1]{Equal contribution}
\endgroup

\begingroup

\footnotetext{Correspondence: \texttt{skodama@middlebury.edu}}
\endgroup

\iclrfinalcopy %
\begin{document}

\maketitle

\begin{abstract}
The rapid spread of text generated by large language models (LLMs) makes it increasingly difficult to distinguish authentic human writing from machine output.
Watermarking offers a promising solution: model owners can embed an imperceptible signal into generated text, marking its origin. Most leading approaches seed an LLM’s next-token sampling with a pseudo-random key that can later be recovered to identify the text as machine-generated, while only minimally altering the model’s output distribution.
However, these methods suffer from two related issues: (i) watermarks are brittle to simple surface-level edits such as paraphrasing or reordering; and (ii) adversaries can append unrelated, potentially harmful text that inherits the watermark, risking reputational damage to model owners. 
To address these issues, we introduce \methodname \footnote{The full code can be found at https://github.com/smid5/SimKey},
a semantic key module that strengthens watermark robustness by tying key generation to the \textit{meaning} of prior context. \methodname uses locality-sensitive hashing over semantic embeddings to ensure that paraphrased text yields the same watermark key, while unrelated or semantically shifted text produces a different one.
Integrated with state-of-the-art watermarking schemes, \methodname improves watermark robustness to paraphrasing and translation while preventing harmful content from false attribution, establishing semantic-aware keying as a practical and extensible watermarking direction.

\end{abstract}

\section{Introduction}

As large language models (LLMs) become widely deployed across various domains, concerns regarding the authenticity and provenance of AI-generated text have grown significantly~\citep{bian2024influence,hanley2024machine,pan2023risk}. Watermarking techniques offer a crucial mechanism for distinguishing between human-authored and machine-generated content~\citep{kuditipudi2024robustdistortionfreewatermarks,yang2023survey}. 
Ideally, a watermarking method should not only provide reliable identification of AI-generated text but also maintain high generation quality. To be practical, watermarks must also be robust against adversarial attempts to remove the watermark or to mark unrelated content.

\begin{figure}
    \centering
    \includegraphics[width=\linewidth]{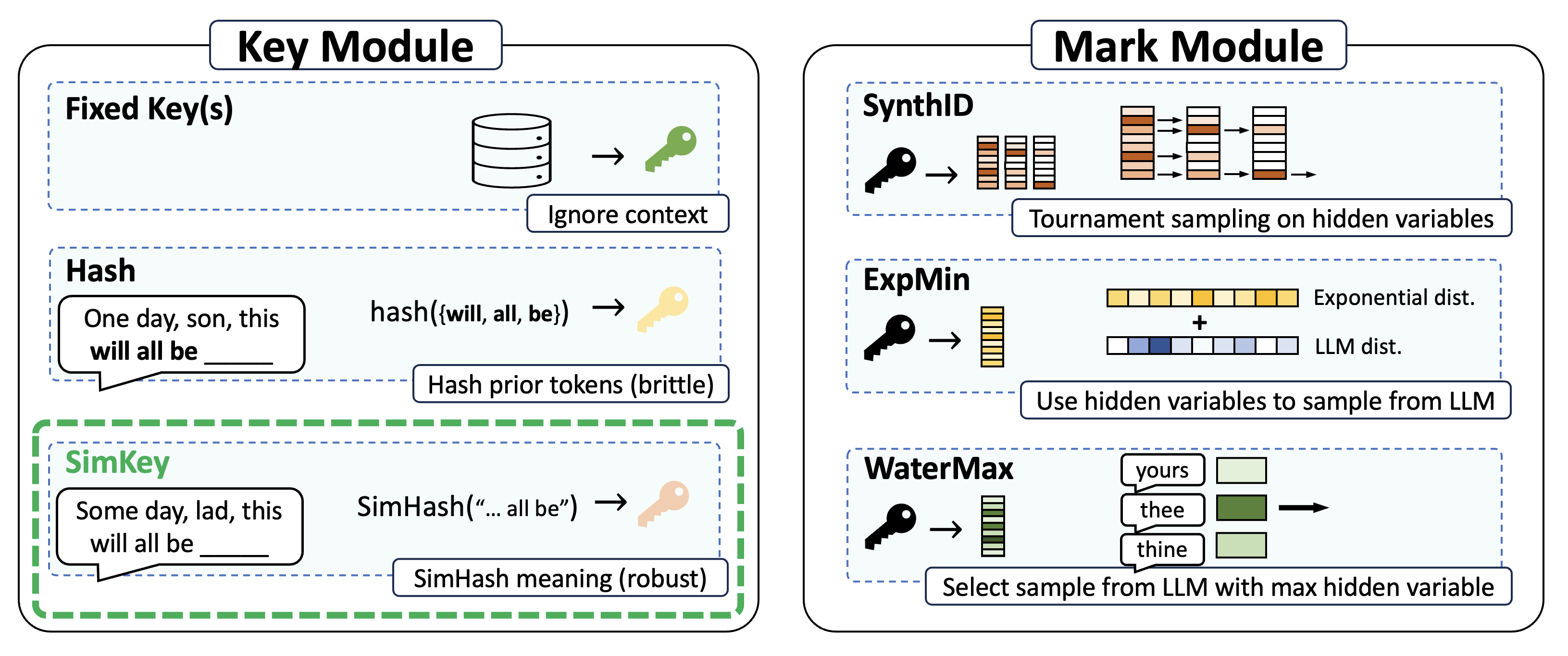}
\caption{\textbf{Common components in watermarking.} The \textit{key module} (left) generates a seed that guides watermarking, using options such as a fixed key (or a fixed set of keys), a hash of prior tokens, or a semantic SimHash of the context (ours). The \textit{mark module} (right) modifies token sampling given the key. E.g., via tournament sampling (SynthID), exponential-min sampling (ExpMin), or selecting generations that maximize a hidden variable (WaterMax).}
    \label{fig:modules}
\end{figure}

These practical considerations make embedding watermarks into generated text inherently challenging. Most methods use a \textit{mark module} that modifies token generation, and a \textit{key module} that conditions the mark module on previously generated text~\cite{huang2024waterpool} (see \Cref{fig:modules}). Early approaches use a mark module that increases the likelihood of certain token sequences (e.g. the now canonical ``red-green'' list~\citep{zhao2023provable,kirchenbauer2024watermarklargelanguagemodels}), but this often introduces fluency-degrading distortions \citep{rastogi2024revisiting}. Moreover, such patterns can be exploited by adversaries~\citep{sadasivan2023can,jovanovic2024watermark}. To mitigate these issues, other methods leverage \textit{pseudo-random next-token selection}. Concretely, they use a secret random variable (i.e. the \textit{key}) to seed the sampling process, which keeps outputs consistent with the LLM distribution \citep{sadasivan2023can,kuditipudi2024robustdistortionfreewatermarks,liu2025survey}.

While watermark \textit{embedding} techniques have been widely studied, \textit{key} generation remains largely unchanged. Keys can be generated through a context-\textit{independent} key module; e.g., by using a cyclic key or sampling from a given key pool. However, reusing keys introduces patterns that undermine both the quality and security of the generated text~\citep{kuditipudi2024robustdistortionfreewatermarks}. Using many different keys reduces these effects, but yields watermarks that are harder to detect and computationally less efficient. A context-\textit{dependent} approach might hash prior tokens to generate keys; however, hashing the last few tokens causes the model to reuse common phrases, and introduce brittleness to small context changes~\citep{kirchenbauer2024watermarklargelanguagemodels}. Additionally, both context-\textit{independent} and -\textit{dependent} approaches risk compromising the watermark owner's reputation, since an adversary can insert harmful text into a watermarked passage, which will still be flagged as model-generated.

To address this, we introduce a key module that uses the locality-sensitive hashing (LSH) technique SimHash \citep{charikar2002similarity}.
By applying SimHash to a semantic embedding of the preceding context, our method ties the key to the \textit{meaning} of the text.
As a key module, \methodname is \textbf{(i) robust to semantic paraphrasing}: when meaning is preserved (i.e., semantic embeddings are similar), the key tends to remain the same.
Simultaneously, \methodname is \textbf{(ii) sensitive to meaning-changing edits}: if watermarked text is moved out of context or if unrelated (potentially harmful) tokens are inserted, the key is likely to change.
Finally, \methodname ensures \textbf{(iii) a sufficiently large and diverse key space}, since \methodname varies the key with semantics and across multiple hash identities.

We emphasize that \methodname is a general key module that can be paired with many different mark modules.
In this work, we demonstrate \methodname combined with three state-of-the-art mark modules: distortion-free exponential minimum sampling (ExpMin) \citep{kuditipudi2024robustdistortionfreewatermarks}, SynthID \citep{Dathathri2024Scalable}, and WaterMax \citep{giboulot2024watermax}. 
We describe how to use \methodname in \Cref{sec:method}. 
In \Cref{sec:results}, we evaluate \methodname and confirm it is sensitive to unrelated (potentially harmful) content insertion while remaining robust to meaning-preserving transformations such as paraphrasing and translation. We end with a discussion of limitations and broader implications.

\begin{figure}
    \centering
    \includegraphics[width=1\linewidth]{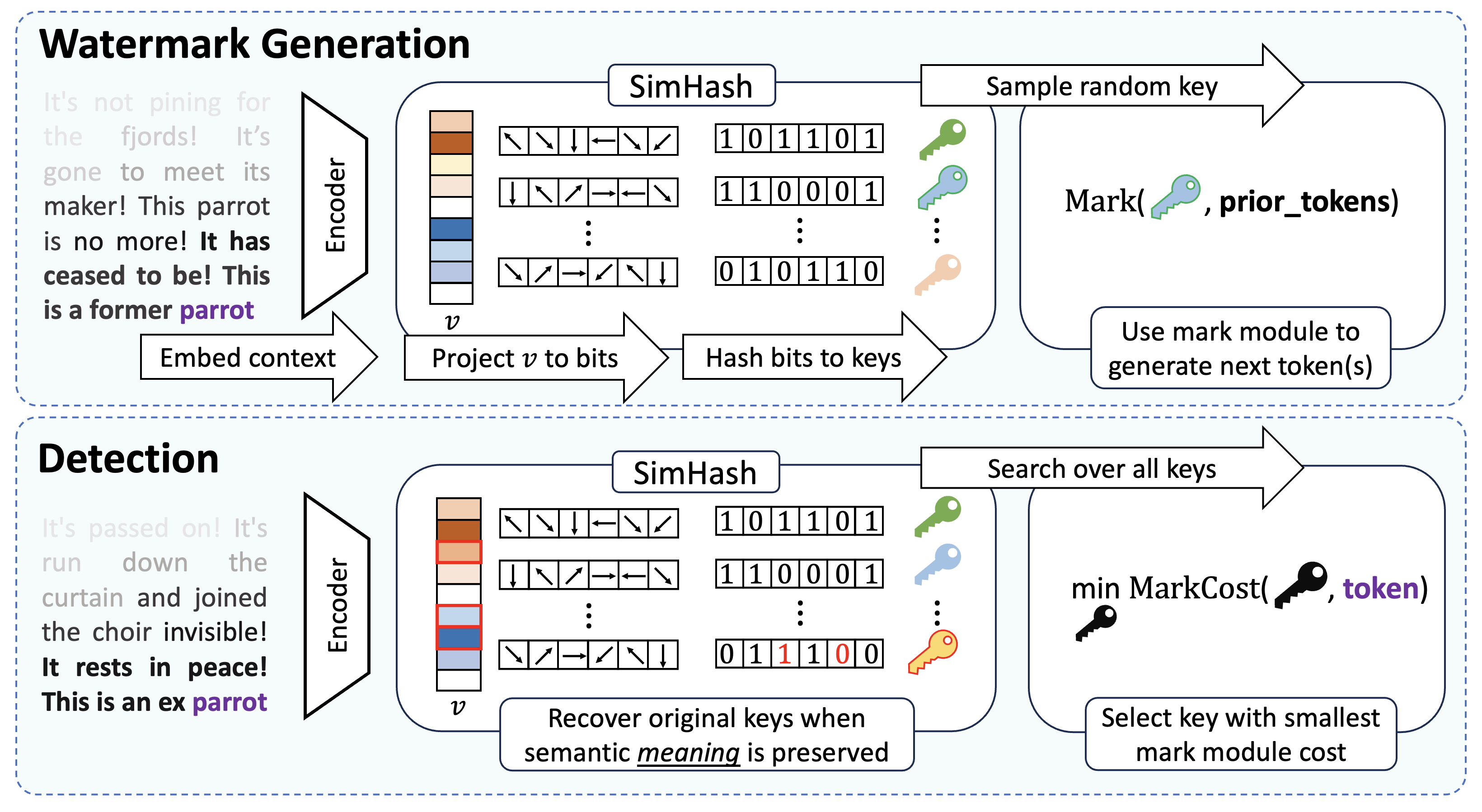}
    \caption{\textbf{Overview of our semantic watermarking method, \methodname.} \textit{Generation (top):} we embed the preceding context into a semantic vector $v$, project onto random directions, and take signs (SimHash), then hash the resulting bits to seed keys that modulate the LLM sampling (e.g., Gumbel/ExpMin sampling). \textit{Detection (bottom):} we re-embed the context before each token, recompute \methodname, and use the mark module’s alignment cost to select the best-matching key per position. }
    \label{fig:simmark_overview}
\end{figure}

\vspace{-1em}
\section{Preliminaries}
\label{sec:background}
\begin{fancybox}{SimHash}
Originally developed for efficient approximate nearest neighbor search, locality sensitive hashing (LSH) provides embeddings that preserve similarity \citep{indyk1998approximate, gionis1999similarity}. 
SimHash \citep{charikar2002similarity} is one such LSH approach that embeds an input vector by random projections so that similar inputs yield similar bit patterns. 
The benefit of SimHash over standard hashing is that nearby vectors $\mathbf{v}$ and $\mathbf{v'}$ are more likely to agree on bits, with agreement controlled by the angle between them:
\begin{align}
    \theta(\mathbf{v}, \mathbf{v'}) = \arccos( \frac{\langle\mathbf{v}, \mathbf{v}'\rangle}{\| \mathbf{v} \|_2 \| \mathbf{v}'\|_2}).
    \label{eq:angle}
\end{align}
\end{fancybox}

To implement SimHash, we choose $b$ random unit vectors $\{\mathbf{r}_j\}_{j=1}^b$, project the vector $\mathbf{v}$ onto each, and record the sign to produce a $b$-bit sequence, which we then hash to obtain a pseudo-random output (\Cref{alg:simhash}). The probability of reproducing the same key for two semantic embeddings $\mathbf{v}$ and $\mathbf{v'}$ is then given as a function of the angle $\theta(\mathbf{v}, \mathbf{v}')$ in Lemma~\ref{lemma:similar}.

\begin{lemma}[SimHash Guarantee \citep{charikar2002similarity}]\label{lemma:similar}
    Consider two vectors $\mathbf{v}$ and $\mathbf{v}'$, with angle $\theta(\mathbf{v}, \mathbf{v}')$.
    For a fixed input (i.e., the same secret salt and key index), Algorithm \ref{alg:simhash} produces the same key with probability~,
        $\left(1- \frac{\theta(v,v')}{180^{\circ}}\right)^b~.$
\end{lemma}

Increasing $b$ will decrease the probability of a match in the key, especially for far apart vectors with low angle.

\begin{algorithm}[b!]
    \caption{\methodname}
    \label{alg:simhash}
    \begin{algorithmic}
    \STATE \textbf{Input:} $\mathbf{v}$: semantic vector, \texttt{idx}: key index, \texttt{salt}: secret salt, $b$: number of bits, \texttt{hash}: cryptographic hash function
    \STATE \textbf{Output:} Semantically and securely generated key
    \STATE $\texttt{bits} \gets \mathbf{0}$ \hfill $\triangleright$ Initialize hash input 
    \FOR{$j = 1, \dots, b$}
        
        \STATE $s \gets \texttt{hash}(\texttt{idx}, j, \texttt{salt})$
        \STATE Sample $\mathbf{r}_j \overset{s}{\sim} \mathcal{N}(\mathbf{0}, \mathbf{I})$  \hfill $\triangleright$ Reproducibly sample random projection vector \textit{with} key index
        \STATE $\texttt{bits}[j] \gets \text{sign}(\langle \mathbf{v}, \mathbf{r}_j \rangle)$ \hfill $\triangleright$ Random projection
    \ENDFOR
    \STATE \texttt{key} $\gets$ $\texttt{hash}(\texttt{bits}, \texttt{idx}, \texttt{salt})$
    \STATE \textbf{return} \texttt{key}
   \end{algorithmic}
\end{algorithm}

\paragraph{Mark Modules} The \textit{mark module} is the part of any existing watermarking technique that modifies the next token generation of the underlying LLM.
\methodname is flexible and compatible with most existing mark modules. To apply \methodname to existing methods, we need only assume their mark module provides two functions: 
\begin{itemize}
    \item \texttt{Mark}: maps a random key (provided by \methodname) and the prior tokens to the next token(s), using an LLM, and any internal watermarking logic.
    \item \texttt{MarkCost}: maps a key and a token to an \textit{alignment cost}, a real number measuring the likelihood that the generated token was produced by \texttt{Mark}, given a candidate key. Without loss of generality, we assume a lower cost indicates a higher likelihood.
\end{itemize}

\section{\methodname - Semantic and Distortion Free Watermarking}
\label{sec:method}

Our goal is to attach a watermark to the \textit{meaning} of text rather than to \textit{exact} token sequences. This serves two purposes: we want watermarks to \textit{persist} when text is paraphrased to obscure origin, \textit{and} we would like the watermark to \textit{disappear} if unrelated, potentially harmful content is added.

Existing watermarking methods often fail to achieve this because their detection depends on the exact sequence of preceding tokens rather than their meaning. This makes them vulnerable to removal attacks, where even minor rewordings can erase the watermark. Conversely, \methodname,  computes a semantic embedding of prior context and applies SimHash to produce the key. Importantly, \methodname is not itself a new watermarking scheme. Instead, it is a general and flexible component that can augment existing schemes, improving their robustness. To demonstrate how it works with existing \texttt{Mark} modules, we integrate \methodname into (1) distortion-free exponential minimum sampling (ExpMin) \citep{kuditipudi2024robustdistortionfreewatermarks}, (2) tournament-style sampling (SynthID) \citep{Dathathri2024Scalable}, and (3) max-normal style sampling (WaterMax) \cite{giboulot2024watermax}. 
We describe the creation and detection procedures in the following subsections and provide pseudocode in Algorithms \ref{alg:generation} and \ref{alg:detection}.

\begin{algorithm}[bt]
    \caption{Generation with \methodname}
    \label{alg:generation}
    \begin{algorithmic}
    \STATE \textbf{Input:} \texttt{Mark}: mark module for generating next tokens from a key, \texttt{tokens}: prior tokens, $V$: vocabulary size, $k$: number of used hash function identities, $b$: number of bits, \texttt{salt}: secret salt
    \STATE \textbf{Output:} Watermarked token drawn from LLM distribution
    \STATE $\mathbf{v} \gets \text{Embed}(\texttt{tokens})$ \hfill $\triangleright$ Semantically embed prior tokens
    \STATE $\texttt{idx} \sim \text{Uniform}(\{1,\ldots,k\})$ \hfill $\triangleright$ Randomly select key index
    \STATE $\texttt{key} \gets \methodname(\mathbf{v}, \texttt{idx}, \texttt{salt})$ (i.e. \Cref{alg:simhash})
    \STATE $\texttt{next\_token} \gets \texttt{Mark}(\texttt{key},\texttt{tokens})$ 
    \hfill $\triangleright$  Sample next token using watermark method and key%
    \STATE \textbf{return} \texttt{next\_token}
   \end{algorithmic}
\end{algorithm}

\subsection{Key Generation}
\label{subsec:key_generation}
Our goal with \methodname is to attach a key to \textit{what the context means}, not merely to \textit{what the last few tokens were}. Concretely, before each generation step, \methodname embeds the prior context into a semantic vector $\mathbf{v}$ that captures its meaning\footnote{Recent advances in language modeling have made powerful semantic embedders abundant; in this paper we use \texttt{all-MiniLM-L6-v2} \citep{reimers2019sentence}, a sentence-transformer model that encodes input text into 384-dimensional embeddings, although many other similar models are available.}. \methodname then applies SimHash (\Cref{eq:angle}) to convert that vector into a compact, reproducible bit pattern: we project $\mathbf{v}$ onto $b$ random directions and encode the signs of the projections as bits: 
\begin{align*}
    \texttt{bits} = \left[\text{sign}(r_1^\top v), \text{sign}(r_2^\top v), \dots, \text{sign}(r_b^\top v)\right] \in \{-1, 1\}^b~,
\end{align*}
Next, we use a cryptographic hash to produce the \texttt{key}, which we pass to \texttt{Mark} module to sample the next token from the underlying watermarking scheme. The full procedure is given in Algorithm~\ref{alg:simhash}

\textbf{Key index variation.}
In long generations the same semantic state can reappear (following similar context embeddings), which risks reusing identical keys too often.  We therefore randomly draw an index \texttt{idx} from $\{1,\dots,k\}$ at each step, effectively selecting among $k$ independent SimHash instances within \methodname. This maintains semantic stability, in that as long as the meaning remains similar, the right key can still be recovered for \textit{some} index. At the same time it reduces repetitive key reuse that could harm fluency.

\begin{algorithm}[tb]
    \caption{\methodname Detection}
    \label{alg:detection}
    \begin{algorithmic}
    \STATE \textbf{Input:} \texttt{tokens}: (possibly) watermarked tokens, \texttt{MarkCost}: a mark module specific function for computing the alignment cost between the tokens and a key, $V$: vocabulary size, $k$: number of keys, $b$: number of bits, \texttt{salt}: secret salt
    \STATE \textbf{Output:} $p$-value: Probability of observing tokens if they were \textit{not} watermarked
    \STATE $\texttt{cost} \gets 0$
    \FOR{$i = 1,\ldots, |\texttt{tokens}|$}
        \STATE \texttt{prior\_tokens} $\gets \{\texttt{tokens}_1, \ldots, \texttt{tokens}_{i-1}\})$
        \STATE $\mathbf{v} \gets \text{Embed}(\texttt{prior\_tokens})$ \hfill $\triangleright$ Semantically embed prior tokens
        \STATE $\texttt{cost}_i \gets \infty$
        \STATE $\triangleright$ Check each key index
        \FOR{$\texttt{idx}=1,\ldots,k$} 
            \STATE $\texttt{key} \gets \methodname(\mathbf{v}, \texttt{idx}, \texttt{salt})$ \hfill $\triangleright$ (i.e. \Cref{alg:simhash})
            \STATE $\texttt{cand\_cost}_\texttt{idx} \gets \texttt{MarkCost}(\texttt{key}, \texttt{prior\_tokens}) $ \hfill $\triangleright$ Candidate cost of instance \texttt{idx}
            \STATE $\texttt{cost}_i \gets \min(\texttt{cost}_i, \texttt{cand\_cost}_\texttt{idx})$
        \ENDFOR
        \STATE $\texttt{cost} \gets \texttt{cost} + \texttt{cost}_i$
    \ENDFOR
    \STATE $p$-value $\gets \Pr( \text{observing a value as large as }\texttt{cost})$ \hfill $\triangleright$ Depends on alignment cost distribution
    \STATE \textbf{return} $p$-value
   \end{algorithmic}
\end{algorithm}

\begin{figure}[b!]
    \centering
    \includegraphics[width=\linewidth]{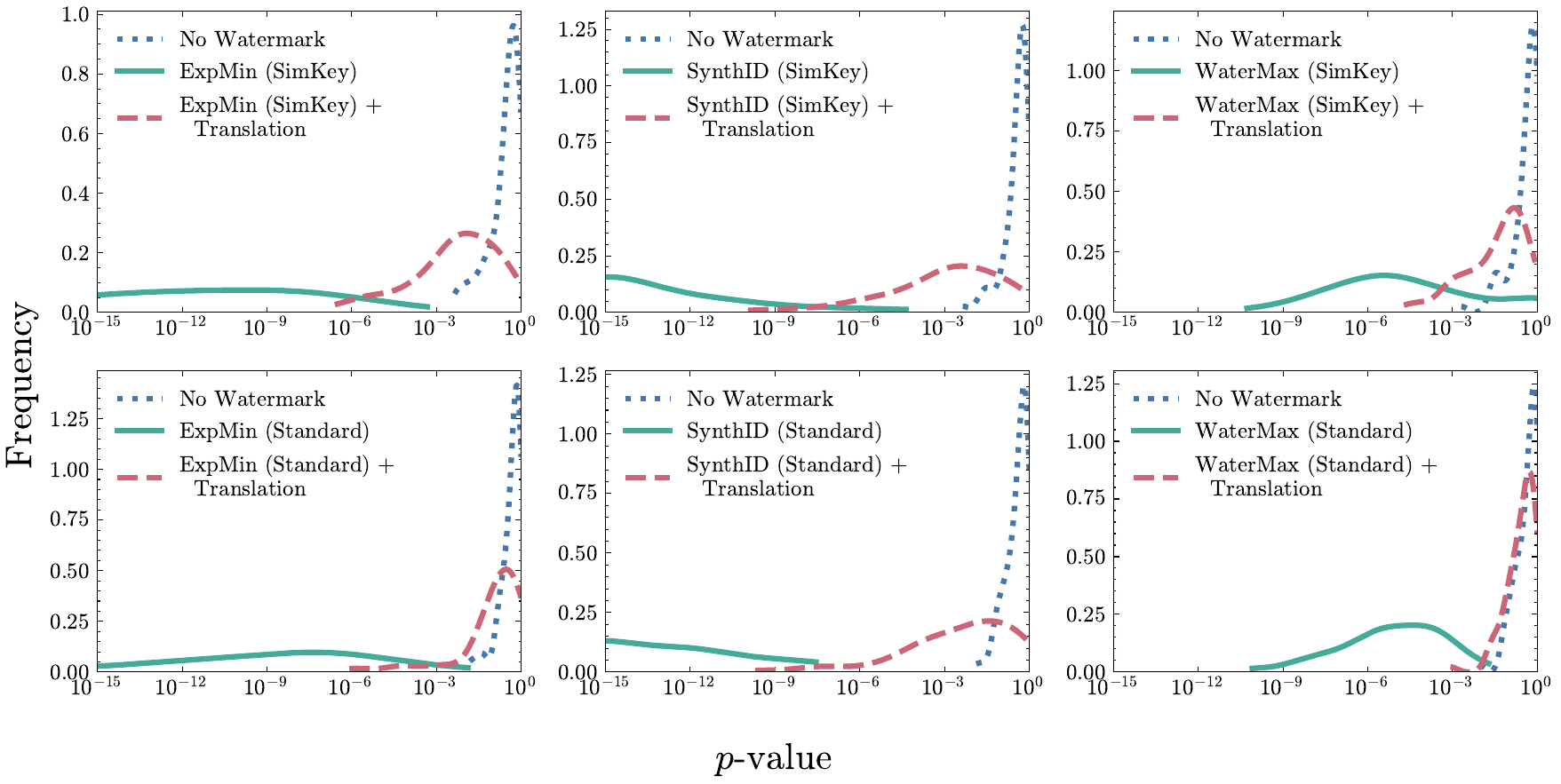}
    \caption{\textbf{Per-token watermark detectability with \methodname and standard hashing}. The $p$-value distributions by mark module (columns) and key module (rows). Shown from $10^0$ to $10^{-15}$, values below this are truncated. \methodname gives similar detectability to standard hashing on the original text. However, after watermarked text is translated to a second language and back, \methodname is more robust, since the key depends on the \textit{meaning} of the text rather than a set of precise tokens.}
    \label{fig:pval-dists-2x3}
\end{figure}

\subsection{Watermark Detection} \label{sec:detection}
During the detection phase, our goal is to recover the same key in order to determine whether the text was likely generated using the watermark. To recover the key, for each position $i$, we re-embed the preceding context to obtain $\mathbf{v}'$ and run \methodname across all $k$ indices to reconstruct candidate keys. Because the text may have been manipulated between generation and detection, $\mathbf{v}'$ may differ from the original context embedding $\mathbf{v}$. Nonetheless, because SimHash depends on \textit{semantic} similarity rather than exact token identity, edits that largely preserve meaning are likely to yield the same key during detection as during generation (the formal probability guarantee is given by Lemma~\ref{lemma:similar}).

Still, even if the watermarked text is unchanged, we do not know which key index was used during generation. To address this, we evaluate all $k$ candidate key indices, where each one is associated with an \textit{alignment cost} noted as $\texttt{cand\_cost}_\texttt{idx}$, and select the one yielding the minimum cost (as defined by the mark module).
The per-token minimum costs are then summed across the entire sequence. We formally describe the procedure in \Cref{alg:detection}.

Finally, to assess whether the resulting total cost provides sufficient evidence of watermarking, we compute a $p$-value: \textbf{the probability of observing a cost at least this low under the null hypothesis} that the text was \textit{not} generated with a watermarking scheme. The formal calculation is given below.

\begin{fancybox}{$p$-value Computation}
Let $\mathcal{D}$ be the distribution of the alignment cost when $\texttt{key}$ was \textit{not} used to generate the tokens.
We assume here that $\mathcal{D}$ is a discrete distribution.
Let $F: \text{supp}(\mathcal{D}) \to [0,1]$ be the CDF of $\mathcal{D}$.
The CDF of the minimum of $k$ independent draws from $\mathcal{D}$ is given by:
\begin{align}
    1-F_{\texttt{cost}}^1(y)
    &= \Pr(\min_{\texttt{idx} \in \{1, \ldots, k\}}
    \texttt{cand\_cost}_\texttt{idx} > y)
   = \prod_{\texttt{idx}=1}^k (1-F(y)) = (1-F(y))^k
\end{align}
where the second equality follows by independence.
\par\vspace{1ex}
We now define the CDF of the sum of $\ell$ independent $\texttt{cost}_i$ samples.
For $\ell > 1$, we recursively define:
\begin{align}
    F_{\texttt{cost}}^\ell (y)
    = \sum_{z \in \text{supp}(\mathcal{D})}
    F_\texttt{cost}^{\ell-1}(z)
    F_\texttt{cost}^{1}(y-z)
    = \text{Convolve}(F_\texttt{cost}^{\ell-1}, F_\texttt{cost}^{1})~.
\end{align}
Finally, we can compute the $p$-value via
\begin{align}
    F_\texttt{cost}^{|\texttt{tokens}|}(\texttt{cost})~.
\end{align}
\end{fancybox}

When the alignment-cost distribution is continuous, the $p$-value can be computed either by discretizing the support or, in some cases, in closed form. For instance, under ExpMin, the alignment cost follows an exponential distribution, yielding a closed form $p$-value (see Appendix~\ref{app:expmin_pvalue} for details).

\section{Results}
\label{sec:results}

\paragraph{Experimental Setup.} 
We conduct all experiments using HuggingFace's \texttt{transformers} library and implement watermarking methods through the \texttt{LogitsProcessor} interface. For text generation, we use top-$p$ sampling with $p=0.9$ to maintain comparable diversity across methods.

\begin{figure}
    \centering
    \includegraphics[width=\linewidth]{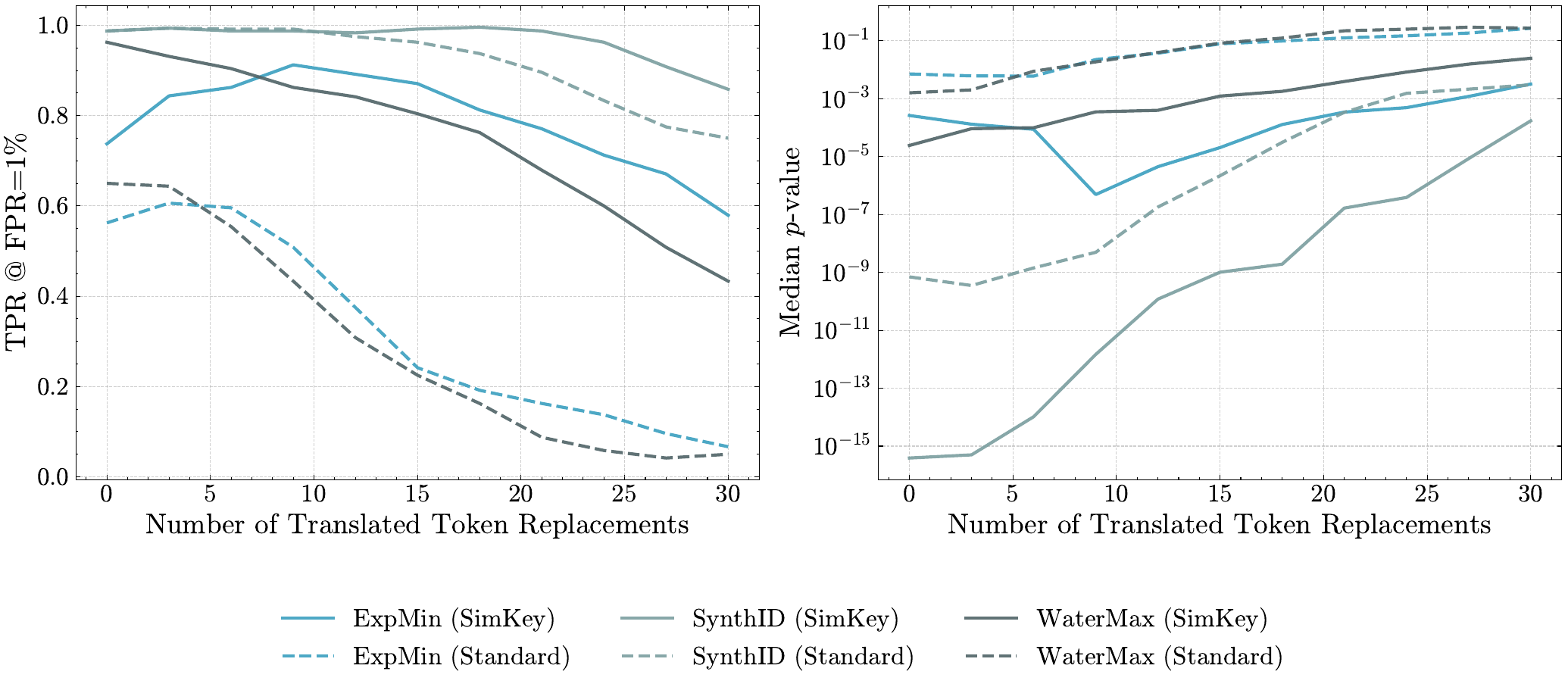}
    \caption{\textbf{\methodname substantially improves detectability under translated token replacements.} At a fixed false positive rate of 1\%, \methodname substantially improves the true positive rate for ExpMin and WaterMax. At first glance, it may appear that SimKey is less effective for SynthID; however, SimKey actually reduces the median $p$-value of SynthID by \textit{several orders of magnitude}. %
    }
    \label{fig:translation}
\end{figure}

\begin{figure}
    \centering
    \includegraphics[width=\linewidth]{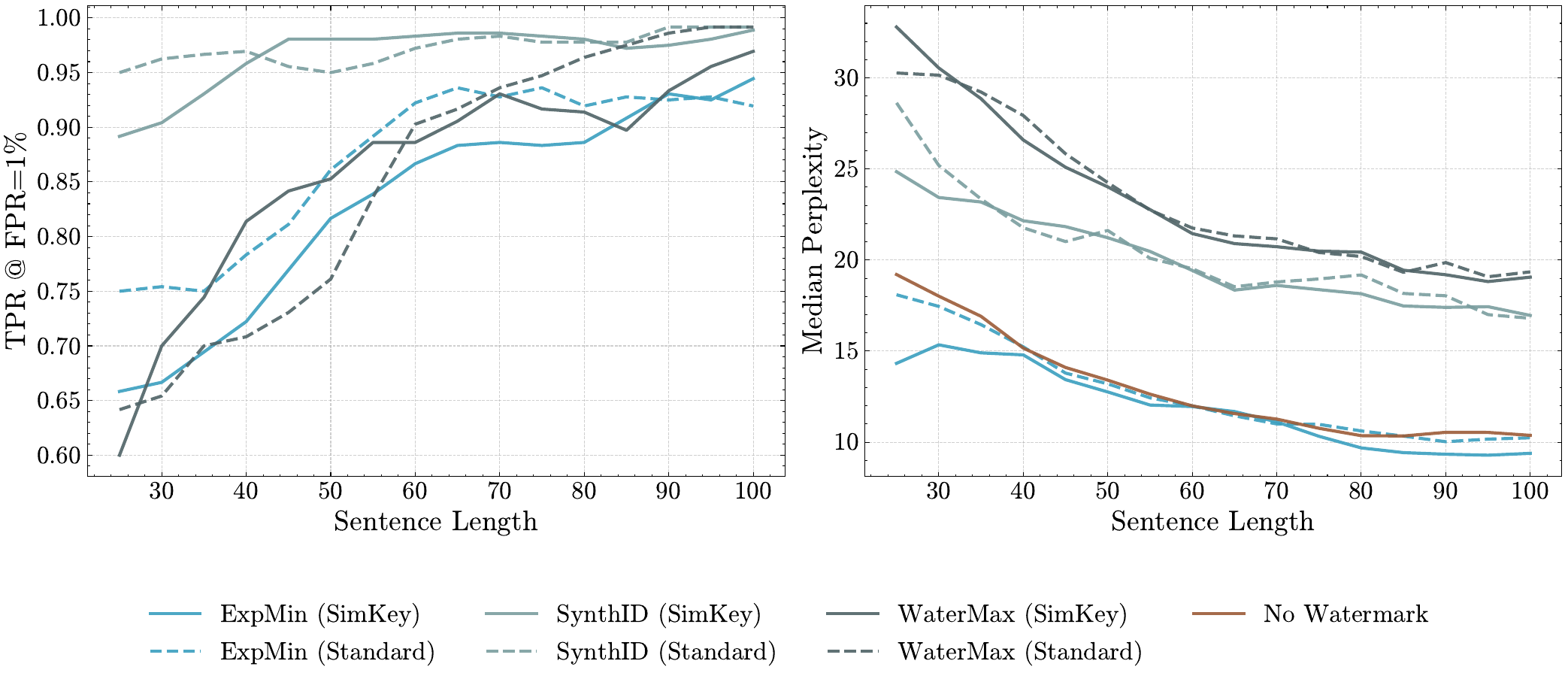}
    \caption{\textbf{\methodname preserves detectability for unmodified text, and the distribution of watermarked text}. (a) Detectability increases with sentence length for all mark modules, with \methodname and standard hashing performing similarly. (b) Perplexity depends on mark module: ExpMin is nearly indistinguishable from unwatermarked text, SynthID has higher perplexity, and WaterMax has the highest; with \methodname and standard hashing performing basically the same.}
    \label{fig:detectability-distortion}
\end{figure}

\textbf{\textit{Base Model.}} For the experiments reported in the main text, we use the quantized version of the Meta Llama 3.1 instruction-tuned 70B model
\texttt{hugging-quants/Meta-Llama-3.1-70B-Instruct-AWQ-INT4}
and the Meta Llama 3 8B model
\texttt{meta-llama/Meta-Llama-3-8B}.
We employ the 70B model for Figure~\ref{fig:detectability-distortion}, and Table~\ref{tab:tpr-token-repl}, and the 8B model for Figures~\ref{fig:pval-dists-2x3} and \ref{fig:translation} due to computational constraints.

\textbf{\textit{Prompt Initialization.}}
Prompts are sampled by drawing three random words to form a short phrase, which serves as a neutral starting point for generation. This procedure avoids strong topical bias while ensuring syntactically valid completions.

\textbf{\textit{Watermarking Parameters.}}
For a fair comparison across all methods, we set the number of keys $k=4$ and the number of bits $b=4$ for the parameters of \methodname. We also set the context window as $8$ tokens for all methods and key modules. 

\textbf{\textit{Perturbations and Attacks.}}
To evaluate robustness, we apply several classes of meaning-preserving and meaning-altering perturbations, which are standard attacks from the watermarking literature \citep{liu2024survey}. The first such attack is the \textbf{Translation Attack}, which involves translating the generated text from English to French and then back to English (we use the \texttt{opus-mt-tc-big-en-fr} and \texttt{opus-mt-tc-big-fr-en} translation models \citep{junczys2018marian}). Translation preserves the overall semantic meaning of the text while perturbing the surface form, sometimes drastically. We also apply \textbf{Translated Token Substitution}, randomly translating individual tokens to French and back, which yields subtle lexical variations without changing either sentence length or global meaning. For the \textbf{Unrelated Token Substitutions} attack, we randomly replace selected token positions with uniformly sampled vocabulary IDs, preserving sequence length but disrupting the semantic meaning. Conversely, for the \textbf{Related Token Substitution} perturbation, we randomly mask tokens and use the BERT base model (cased) \texttt{google-bert/bert-base-cased} to choose the most probable replacement. We repeat this process until a replacement different from the original token is obtained, ensuring that the substituted word remains contextually plausible while subtly altering the surface form.

\paragraph{Summary.} We evaluate \methodname as a key module paired with three mark modules: ExpMin~\citep{kuditipudi2024robustdistortionfreewatermarks}, SynthID~\citep{Dathathri2024Scalable}, and WaterMax~\citep{giboulot2024watermax}, with a standard (non-semantic) hashing key as the baseline. 
Here, we briefly summarize our results before offering a closer analysis: \textbf{(1)} On clean text, \methodname matches the $p$-value distribution of standard hashing (Fig.~\ref{fig:pval-dists-2x3}); \textbf{(2)} Under meaning-preserving edits (paraphrase/translation), \methodname substantially improves detectability (Fig.~\ref{fig:translation}, Table~\ref{tab:tpr-token-repl}); \textbf{(3)} Under meaning-changing edits (unrelated insertions/replacements), \methodname degrades similarly to standard hashing (Table~\ref{tab:tpr-token-repl}), as desired; and \textbf{(4)} Perplexity/distortion is essentially unchanged with respect to standard hashing, with differences dominated by the choice of mark module (Fig.~\ref{fig:detectability-distortion}).

\textbf{\textit{(1) Parity with Standard Hashing.}} When the context is unedited, both key modules recover the correct keys step by step. Consequently, the $p$-value distributions align cleanly across mark modules, as demonstrated in Fig.~\ref{fig:pval-dists-2x3}. This parity is important to practical deployments; \methodname does \textit{not} weaken detection or inflate false positives in benign settings.

\textbf{\textit{(2) Better Detectability Under Paraphrase Edits.}} We find that \methodname is robust to meaning-preserving edits that alter surface forms while keeping semantics close to the original intent. Standard hashing, on the other hand, ties keys to exact recent tokens and thus loses detectability quickly. We can see this effect examining the \textit{translated token replacement} attacks presented in Fig.~\ref{fig:translation}. We observe consistent TPR@1\%FPR gains and lower $p$-values across each of the three mark modules. \textsc{SimKey}’s TPR gains appear mainly at higher token replacement levels where the attack is stronger.

We also show improved detectability with \methodname on entire text segments under the translation attack in \Cref{tab:tpr-translation}, across all three mark modules.

\textbf{\textit{(3) Edits That Change Meaning Remove Watermark.}} For unrelated replacements or insertions that shift topic or intent, watermarks using both \methodname and the baseline key degrade similarly, as evidenced by Table~\ref{tab:tpr-token-repl}. This is the intended behavior of a semantically aware watermark; when semantics drift, the key should change, preventing harmful or off-topic additions from inheriting machine-generated attribution. Additionally, Table~\ref{tab:tpr-token-repl} shows that under related token replacements where the meaning of the text stays the same, \methodname performs \textit{much} better.

\textbf{\textit{(4) Does Not Add Distortion to the  Sampling.}} \methodname only replaces the key module. Because the mark module and base LM distribution are unchanged, the perplexity of the sampled text tracks the standard hashing almost exactly (Fig.~\ref{fig:detectability-distortion}). The primary driver of distortion remains the mark module itself: ExpMin is closest to unwatermarked text, SynthID induces some moderate change in perplexity, and WaterMax induces the greatest change. Using \methodname minimally affects the perplexity or this relative ordering, as expected.

\begin{table}[b!]
\centering
\small
\caption{\textbf{True Positive Rate at Fixed False Positive Rate under different transformations (TPR@FPR$\leq$1\%).} For each method we perturb the text by changing tokens to related or unrelated tokens (under two settings, 15 and 30 modifications). We examine each attack with the standard hashing method (St. Hash) or our method (\methodname). Across both 15 and 30 token replacements we find that \methodname is more robust to related token replacements while being similarly sensitive to the baseline standard hashing scheme to unrelated token replacements.}
\begin{tabular}{l *{8}{S}}
\toprule
& \multicolumn{4}{c}{Unrelated Attack} & \multicolumn{4}{c}{Related Attack} \\
\cmidrule(lr){2-5}\cmidrule(lr){6-9}
Method &
\multicolumn{2}{c}{15 Tokens} & \multicolumn{2}{c}{30 Tokens} &
\multicolumn{2}{c}{15 Tokens} & \multicolumn{2}{c}{30 Tokens} \\
& {St. Hash} & {SimKey} & {St. Hash} & {SimKey} & {St. Hash} & {SimKey} & {St. Hash} & {SimKey} \\
\midrule
ExpMin   & 0.063 & \textbf{0.075} & 0.013 & \textbf{0.038} & 0.025 & \textbf{0.450} & 0.063 & \textbf{0.325} \\
SynthID  &\textbf{0.813} & 0.700 & \textbf{0.225} & 0.138 & 0.500 & \textbf{0.875} & 0.288 & \textbf{0.800} \\
WaterMax & 0.075 & \textbf{0.250} & 0.013 & \textbf{0.050} & 0.025 & \textbf{0.613} & 0.013 & \textbf{0.375} \\
\bottomrule
\end{tabular}
\label{tab:tpr-token-repl}
\end{table}

\section{Limitations}
\label{sec:limitations}

\textbf{Semantic Robustness for the Mark Module.}
While our key module is responsible for determining the seed encodes the semantics of the text, the mark modules, which embeds the watermark into individual tokens, are not semantic in nature. That is, replacing a watermarked token with a synonymous alternative may significantly affect the likelihood of detecting the watermark for that token, even if the seed used for that token generation is correctly recovered. However, in sufficiently long text generations, it is likely that some words or tokens will remain unchanged after transformations such as translation to another language and back. This is especially true for named entities such as people or places, or for punctuation marks, which will often be mapped back to their original token.

A natural extension to our work would be to introduce semantic awareness into the \textit{mark module} as well. For example, one could design key-dependent preferences that favor semantically related words rather than specific surface forms. This would likely further improve the robustness of the watermarking scheme to synonym substitutions and similar removal attacks that we tested in this paper. However, such modifications would trade off with the distortion-free guarantees offered by existing approaches, which is a highly desirable property (see e.g., \Cref{app:exp_samp}, \citep{kuditipudi2024robustdistortionfreewatermarks}). We leave an investigation of semantic mark modules to future work.

\textbf{Utilizing Additional Mark Modules}. We find \methodname to be compatible with most state-of-the-art watermarking techniques. Yet, certain mark modules may require additional adaptation. For example, ``red-green'' list approaches often adjust token probabilities by a fixed shift based on a hash of prior context~\citep{kirchenbauer2024watermarklargelanguagemodels}. When combined with the \textit{key index variation} described in \Cref{subsec:key_generation}, this can lead to unintended behavior: many tokens may appear on the green list for at least one index, \textit{even} in unwatermarked text, thereby weakening detection. In such cases, \methodname can still be applied by disabling index variation, or with potentially other adaptations. More generally, we expect the method to extend naturally to a wide range of existing and future watermarking schemes.

\section{Related Works}

\textbf{Watermarking Language Models.} Most LLM watermarking either perturbs the sampling distribution to embed detectable signals or aims to preserve it while enabling detection. \citet{kirchenbauer2024watermarklargelanguagemodels} partition tokens at each step into pseudo-random ``green''/``red'' sets via a hash of prior tokens. Unigram \citep{zhao2023provablerobustwatermarkingaigenerated} fixes these lists for robustness, but such approaches face spoofing risks \citep{liu2025survey,jovanovic2024watermark,sadasivan2023can}. Related methods include GumbelSoft \citep{fu2024gumbelsoftdiversifiedlanguagemodel}, Duwak \citep{zhu2024duwakdualwatermarkslarge}, SWEET \citep{lee2024wrotecodewatermarkingcode}, and NS-Watermark \citep{takezawa2025necessarysufficientwatermarklarge}. SynthID-text \citep{Dathathri2024Scalable} similarly adjusts sampling to preserve quality and latency. See \citet{liu2025survey} for a comprehensive review.

\textbf{Distortion-Free LLM Watermarking.} \citet{aaronson2023reform} augment the exponential mechanism with a hash of prior tokens. \citet{christ2023undetectablewatermarkslanguagemodels} use cryptographic indistinguishability, making detection without a key computationally hard. Exponential Minimum Sampling (ExpMin) \citep{kuditipudi2024robustdistortionfreewatermarks} seeds Gumbel-Softmax with a pseudo-random sequence, leaving the per-token distribution unchanged; detection then correlates text with the sequence, but removal attacks remain a challenge. Our module replaces fixed seeds (e.g., token-hash or PRNG) with a dynamic key derived via semantic hashing of context, thus addressing a major challenge framed by the authors of ExpMin.

\textbf{Semantic Watermarking.} To resist surface edits, semantic methods embed signals tied to meaning. \citet{liu2024semanticinvariantrobustwatermark} use an external encoder (e.g., BERT) and a learned watermark head, but require training. Other semantic approaches include Remark-LLM \citep{zhang2024remarkllmrobustefficientwatermarking}, SemStamp \citep{hou2024semstampsemanticwatermarkparaphrastic}, and $k$-SemStamp \citep{hou2024ksemstampclusteringbasedsemanticwatermark}. Our approach follows this direction without model-level changes: it derives a local semantic key from context to enable semantic awareness, preserve analytical tractability, and remain compatible with state-of-the-art mark modules \citep{huang2024waterpool}; this is desirable in that \methodname can be readily integrated into existing state of the art watermarking schemes without heavy adaptations.

\section{Conclusion}

We present \methodname, a key module that utilizes Locally Sensitive Hashing to allow embedding of detectable and robust watermarks in language model outputs. 
Our key identity remains stable under edits that preserve semantics, but not under edits that change them.
Through experimental evaluations, we show that our module improves robustness against various attacks while maintaining generation perplexity comparable to the same watermarking methods using existing key modules. These results highlight the potential of semantic-aware keys for watermarking as a practical and principled solution for practical and responsible deployment of language models.

\clearpage

\bibliography{iclr2026_conference}
\bibliographystyle{iclr2026_conference}

\appendix

\clearpage

\section{Exponential Minimum Sampling}
\label{app:exp_samp}

Exponential Minimum Sampling enables randomized token selection based on the LLM’s probabilities in a numerically stable way, while also exhibiting properties that make it effective for watermarking.
Let $\mathbf{p} \in [0,1]^V$ be a distribution over the vocabulary. Suppose $\xi \in [0,1]^V$ is a random variable where each entry is independently drawn from the uniform distribution on $[0,1]$.
Exponential minimum sampling selects the next token via
\begin{align}
\label{eq:expmin}
    i^* \gets \argmin_{i \in \{1,\ldots,|V|\}} \frac{-\log([\xi]_i)}{p_i}.
\end{align}
\begin{lemma}[Exponential Minimum Sampling]
    The probability that a token $i^*$ is selected via exponential minimum sampling in Equation \ref{eq:expmin} is:
    \begin{align*}
    \Pr(i^* \textnormal{ is selected}) =p_{i^*}
    \end{align*}
\end{lemma}
This is a well-known fact that follows from Gumbel sampling see e.g., \cite{kuditipudi2024robustdistortionfreewatermarks}.
For the interested reader, we present a proof below.
\begin{proof}
The first observation is that each term in the minimum is an exponentially distributed random variable with rate $p_i$.
To see this, notice that $-\log(\cdot)$ applied to a uniform variable results in an exponentially distributed random variable with rate $1$, i.e., $- \log([\xi]_i) \sim \text{Exp}(1)$.
Next, observe that dividing an exponentially distributed variable by a constant multiplies its rate by the constant i.e., $\frac{-\log([\xi]_i)}{p_i} \sim \text{Exp}(p_i)$.
We can then directly analyze the probability that a particular $i^*$ achieves the minimum value.
For notational convenience, let $X_i =\frac{-\log([\xi]_i)}{p_i}$.
Then $X_i \sim \text{Exp}(p_i)$ and 
\begin{align*}
\Pr&\left(i^* = \argmin_{i \in \{1,\ldots,|V|\}} \frac{-\log([\xi]_i)}{p_i} \right)
= \int_{x=0}^\infty \Pr(X_{i^*} = x) \Pr(\forall_{i\neq i^*} X_i > x) dx
\\&= \int_{x=0}^\infty p_{i^*} e^{-p_{i^*} x}
\left(\prod_{i \neq i^*} e^{-p_i x} \right) dx
= p_{i^*} \int_{x=0}^\infty e^{-(p_1 + \ldots p_V)x} dx
= p_{i^*}
\end{align*}
where the second equality follows by plugging in the PDF and CDF of the exponential distribution.
The statement immediately follows.
\end{proof}

\subsection{Closed-form $p$-value under ExpMin}
\label{app:expmin_pvalue}

Recall from Section~\ref{sec:detection} that at position $t$ we evaluate all $k$ candidate key indices and take the per-token alignment cost as the minimum across candidates. Under ExpMin, for a fixed token $y_t$ and key index $j$, the quantity,
\begin{align*}
    Z_{t,j} := - \log([\xi_{t,j}]_{y_t})~,
\end{align*}
is exponentially distributed with rate 1. This is because $[\xi_{t,j}]_{y_t} \sim \mathrm{Unif}[0,1]$. Taking the min across $k$ independent candidates then yields,
\begin{align*}
    C_t := \min_{j \in \{1, \ldots k\}} Z_{t,j} \sim \mathrm{Exp}(k)~,
\end{align*}
with $Pr[C_t > c] = e^{-kc}$.

If we assume independence across positions\footnote{This holds exactly if the seeds at different positions are independent; in practice it is an accurate approximation because the per-position seeds are (pseudo)random functions of the preceding context and key index.}, then the total alignment cost over the $n$ tokens is,
\begin{align*}
    S_n := \sum_{t=1}^n C_t~,
\end{align*}
has a Gamma distribution with the shape $n$ and rate $k$ i.e. $S_n \sim \mathrm{Gamma}(\text{shape}=n,~\text{rate}=k)$ \citep{ross2023introduction}. Its CDF admits the closed form,
\begin{align*}
    F_{S_n}(s) = Pr[S_n \leq s] = \frac{\gamma(n, ks)}{\Gamma(n)} = 1 - e^{-ks} \sum_{m=0}^{n-1}\frac{(ks)^m}{m!}~.
\end{align*}
Because lower costs are stronger evidence of watermarking, the one-sided $p$-value is,
\begin{align*}
    p = F_{S_n}(s_{obs}) = 1 - e^{-k s_{obs}} \sum_{m=0}^{n-1} \frac{(k s_{obs})^m}{m!}~.
\end{align*}

However, in our implementation, we report the \textit{mean} cost $\bar{S_n} := S_n/n$ instead of the sum. Since, $\bar{S_n} \sim \mathrm{Gamma}(\text{shape}=n,~\text{rate}=kn)$, the corresponding CDF is,
\begin{align*}
    F_{\bar{S_n}}(a) = Pr[\bar{S_n} \leq a] = \frac{\gamma(n, k n a)}{\Gamma(n)} = 1 - e^{-kna}\sum_{m=0}^{n-1}\frac{(kna)^m}{m!}~,
\end{align*}
which means the $p$-value is $p = F_{\bar{S_n}}(a_{obs})$. This is equivalent to the \textit{sum} up to the deterministic scaling by $n$.

\section{Additional Results}

\subsection{Detectability.}

\begin{figure}[H]
    \centering
    \includegraphics[width=0.7\linewidth]{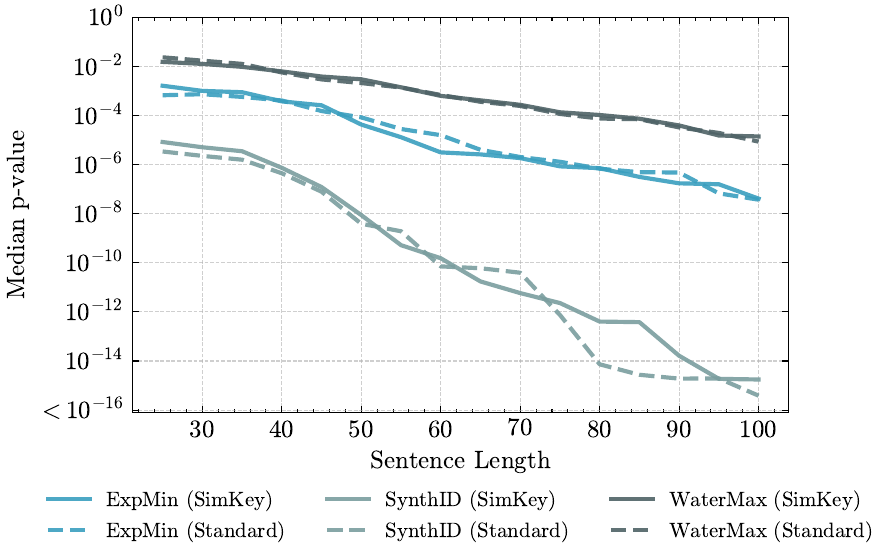}
    \caption{The median $p$-value among 80 generated texts for each sentence length. \methodname preserves the detectability, with the median $p$-value staying basically the same between \methodname and standard hashing.}
    \label{fig:sentence-length-appendix}
\end{figure}

\subsection{Robustness to translation attacks.}

In \Cref{tab:tpr-translation} we report the detectability (True Positive Rate at fixed False Positive rate) for the translation attack described in \Cref{sec:results}, for entire text segments. We generate 120 texts of unwatermarked text and 80 texts of watermarked text that is then translated into French and back. \methodname improve the robustness under translation across all three examined mark modules. 

\begin{table}[b!]
\centering
\small
\caption{\textbf{True Positive Rate under translation attack at FPR$\leq$1\%.}}
\begin{tabular}{lcc}
\toprule
Method & St. Hash & SimKey \\
\midrule
ExpMin   & 0.113 & \textbf{0.500} \\
SynthID  & 0.625 & \textbf{0.688} \\
WaterMax & 0.013 & \textbf{0.300} \\
\bottomrule
\end{tabular}
\label{tab:tpr-translation}
\end{table}

\section{Use of LLMs}
We used LLMs to polish the writing and find typos.

\end{document}

%% file: math_commands.tex
\usepackage{amsmath,amsfonts,bm}

\def\eqref#1{equation~\ref{#1}}

\def\1{\bm{1}}

\DeclareMathAlphabet{\mathsfit}{\encodingdefault}{\sfdefault}{m}{sl}
\SetMathAlphabet{\mathsfit}{bold}{\encodingdefault}{\sfdefault}{bx}{n}

\DeclareMathOperator*{\argmin}{arg\,min}

%% file: arxiv.bbl
\begin{thebibliography}{32}
\providecommand{\natexlab}[1]{#1}
\providecommand{\url}[1]{\texttt{#1}}
\expandafter\ifx\csname urlstyle\endcsname\relax
  \providecommand{\doi}[1]{doi: #1}\else
  \providecommand{\doi}{doi: \begingroup \urlstyle{rm}\Url}\fi

\bibitem[Aaronson(2023)]{aaronson2023reform}
Scott Aaronson.
\newblock reform’ai alignment with scott aaronson.
\newblock \emph{AXRP-the AI X-risk Research Podcast}, 2023.

\bibitem[Bian et~al.(2024)Bian, Lin, Liu, Lu, Zhang, He, Han, and Sun]{bian2024influence}
Ning Bian, Hongyu Lin, Peilin Liu, Yaojie Lu, Chunkang Zhang, Ben He, Xianpei Han, and Le~Sun.
\newblock Influence of external information on large language models mirrors social cognitive patterns.
\newblock \emph{IEEE Transactions on Computational Social Systems}, 2024.

\bibitem[Charikar(2002)]{charikar2002similarity}
Moses~S Charikar.
\newblock Similarity estimation techniques from rounding algorithms.
\newblock In \emph{Proceedings of the thiry-fourth annual ACM symposium on Theory of computing}, pp.\  380--388, 2002.

\bibitem[Christ et~al.(2023)Christ, Gunn, and Zamir]{christ2023undetectablewatermarkslanguagemodels}
Miranda Christ, Sam Gunn, and Or~Zamir.
\newblock Undetectable watermarks for language models, 2023.
\newblock URL \url{https://arxiv.org/abs/2306.09194}.

\bibitem[Dathathri et~al.(2024)Dathathri, See, Ghaisas, et~al.]{Dathathri2024Scalable}
Saurabh Dathathri, Abigail See, Shivani Ghaisas, et~al.
\newblock Scalable watermarking for identifying large language model outputs.
\newblock \emph{Nature}, 634:\penalty0 818--823, 2024.
\newblock \doi{10.1038/s41586-024-08025-4}.
\newblock URL \url{https://doi.org/10.1038/s41586-024-08025-4}.

\bibitem[Fu et~al.(2024)Fu, Zhao, Yang, Zhang, Chen, and Xiao]{fu2024gumbelsoftdiversifiedlanguagemodel}
Jiayi Fu, Xuandong Zhao, Ruihan Yang, Yuansen Zhang, Jiangjie Chen, and Yanghua Xiao.
\newblock Gumbelsoft: Diversified language model watermarking via the gumbelmax-trick, 2024.
\newblock URL \url{https://arxiv.org/abs/2402.12948}.

\bibitem[Giboulot \& Furon(2024)Giboulot and Furon]{giboulot2024watermax}
Eva Giboulot and Teddy Furon.
\newblock Watermax: breaking the llm watermark detectability-robustness-quality trade-off.
\newblock \emph{Advances in Neural Information Processing Systems}, 37:\penalty0 18848--18881, 2024.

\bibitem[Gionis et~al.(1999)Gionis, Indyk, Motwani, et~al.]{gionis1999similarity}
Aristides Gionis, Piotr Indyk, Rajeev Motwani, et~al.
\newblock Similarity search in high dimensions via hashing.
\newblock In \emph{Vldb}, volume~99, pp.\  518--529, 1999.

\bibitem[Hanley \& Durumeric(2024)Hanley and Durumeric]{hanley2024machine}
Hans~WA Hanley and Zakir Durumeric.
\newblock Machine-made media: Monitoring the mobilization of machine-generated articles on misinformation and mainstream news websites.
\newblock In \emph{Proceedings of the International AAAI Conference on Web and Social Media}, volume~18, pp.\  542--556, 2024.

\bibitem[Hou et~al.(2024{\natexlab{a}})Hou, Zhang, He, Wang, Chuang, Wang, Shen, Durme, Khashabi, and Tsvetkov]{hou2024semstampsemanticwatermarkparaphrastic}
Abe~Bohan Hou, Jingyu Zhang, Tianxing He, Yichen Wang, Yung-Sung Chuang, Hongwei Wang, Lingfeng Shen, Benjamin~Van Durme, Daniel Khashabi, and Yulia Tsvetkov.
\newblock Semstamp: A semantic watermark with paraphrastic robustness for text generation, 2024{\natexlab{a}}.
\newblock URL \url{https://arxiv.org/abs/2310.03991}.

\bibitem[Hou et~al.(2024{\natexlab{b}})Hou, Zhang, Wang, Khashabi, and He]{hou2024ksemstampclusteringbasedsemanticwatermark}
Abe~Bohan Hou, Jingyu Zhang, Yichen Wang, Daniel Khashabi, and Tianxing He.
\newblock k-semstamp: A clustering-based semantic watermark for detection of machine-generated text, 2024{\natexlab{b}}.
\newblock URL \url{https://arxiv.org/abs/2402.11399}.

\bibitem[Huang \& Wan(2024)Huang and Wan]{huang2024waterpool}
Baizhou Huang and Xiaojun Wan.
\newblock Waterpool: A watermark mitigating trade-offs among imperceptibility, efficacy and robustness.
\newblock \emph{arXiv preprint arXiv:2405.13517}, 2024.

\bibitem[Indyk \& Motwani(1998)Indyk and Motwani]{indyk1998approximate}
Piotr Indyk and Rajeev Motwani.
\newblock Approximate nearest neighbors: towards removing the curse of dimensionality.
\newblock In \emph{Proceedings of the thirtieth annual ACM symposium on Theory of computing}, pp.\  604--613, 1998.

\bibitem[Jovanovi{\'c} et~al.(2024)Jovanovi{\'c}, Staab, and Vechev]{jovanovic2024watermark}
Nikola Jovanovi{\'c}, Robin Staab, and Martin Vechev.
\newblock Watermark stealing in large language models.
\newblock \emph{arXiv preprint arXiv:2402.19361}, 2024.

\bibitem[Junczys-Dowmunt et~al.(2018)Junczys-Dowmunt, Grundkiewicz, Dwojak, Hoang, Heafield, Neckermann, Seide, Germann, Aji, Bogoychev, et~al.]{junczys2018marian}
Marcin Junczys-Dowmunt, Roman Grundkiewicz, Tomasz Dwojak, Hieu Hoang, Kenneth Heafield, Tom Neckermann, Frank Seide, Ulrich Germann, Alham~Fikri Aji, Nikolay Bogoychev, et~al.
\newblock Marian: Fast neural machine translation in c++.
\newblock In \emph{Proceedings of ACL 2018, System Demonstrations}, pp.\  116--121, 2018.

\bibitem[Kirchenbauer et~al.(2024)Kirchenbauer, Geiping, Wen, Katz, Miers, and Goldstein]{kirchenbauer2024watermarklargelanguagemodels}
John Kirchenbauer, Jonas Geiping, Yuxin Wen, Jonathan Katz, Ian Miers, and Tom Goldstein.
\newblock A watermark for large language models, 2024.
\newblock URL \url{https://arxiv.org/abs/2301.10226}.

\bibitem[Kuditipudi et~al.(2024)Kuditipudi, Thickstun, Hashimoto, and Liang]{kuditipudi2024robustdistortionfreewatermarks}
Rohith Kuditipudi, John Thickstun, Tatsunori Hashimoto, and Percy Liang.
\newblock Robust distortion-free watermarks for language models.
\newblock \url{https://arxiv.org/abs/2307.15593}, 2024.

\bibitem[Lee et~al.(2024)Lee, Hong, Ahn, Hong, Lee, Yun, Shin, and Kim]{lee2024wrotecodewatermarkingcode}
Taehyun Lee, Seokhee Hong, Jaewoo Ahn, Ilgee Hong, Hwaran Lee, Sangdoo Yun, Jamin Shin, and Gunhee Kim.
\newblock Who wrote this code? watermarking for code generation, 2024.
\newblock URL \url{https://arxiv.org/abs/2305.15060}.

\bibitem[Liu et~al.(2024{\natexlab{a}})Liu, Pan, Hu, Meng, and Wen]{liu2024semanticinvariantrobustwatermark}
Aiwei Liu, Leyi Pan, Xuming Hu, Shiao Meng, and Lijie Wen.
\newblock A semantic invariant robust watermark for large language models, 2024{\natexlab{a}}.
\newblock URL \url{https://arxiv.org/abs/2310.06356}.

\bibitem[Liu et~al.(2024{\natexlab{b}})Liu, Pan, Lu, Li, Hu, Zhang, Wen, King, Xiong, and Yu]{liu2024survey}
Aiwei Liu, Leyi Pan, Yijian Lu, Jingjing Li, Xuming Hu, Xi~Zhang, Lijie Wen, Irwin King, Hui Xiong, and Philip Yu.
\newblock A survey of text watermarking in the era of large language models.
\newblock \emph{ACM Computing Surveys}, 57\penalty0 (2):\penalty0 1--36, 2024{\natexlab{b}}.

\bibitem[Liu et~al.(2025)Liu, Pan, Lu, Li, Hu, Zhang, Wen, King, Xiong, and Yu]{liu2025survey}
Aiwei Liu, Leyi Pan, Yijian Lu, Jingjing Li, Xuming Hu, Xi~Zhang, Lijie Wen, Irwin King, Hui Xiong, and Philip Yu.
\newblock A survey of text watermarking in the era of large language models.
\newblock \emph{ACM Computing Surveys}, 57\penalty0 (2):\penalty0 1--36, February 2025.
\newblock \doi{10.1145/3691626}.
\newblock URL \url{https://doi.org/10.1145/3691626}.

\bibitem[Pan et~al.(2023)Pan, Pan, Chen, Nakov, Kan, and Wang]{pan2023risk}
Yikang Pan, Liangming Pan, Wenhu Chen, Preslav Nakov, Min-Yen Kan, and William~Yang Wang.
\newblock On the risk of misinformation pollution with large language models.
\newblock \emph{arXiv preprint arXiv:2305.13661}, 2023.

\bibitem[Rastogi \& Pruthi(2024)Rastogi and Pruthi]{rastogi2024revisiting}
Saksham Rastogi and Danish Pruthi.
\newblock Revisiting the robustness of watermarking to paraphrasing attacks.
\newblock In \emph{Proceedings of the 2024 Conference on Empirical Methods in Natural Language Processing}, pp.\  18100--18110, 2024.

\bibitem[Reimers \& Gurevych(2019)Reimers and Gurevych]{reimers2019sentence}
Nils Reimers and Iryna Gurevych.
\newblock Sentence-bert: Sentence embeddings using siamese bert-networks.
\newblock In \emph{Proceedings of the 2019 Conference on Empirical Methods in Natural Language Processing and the 9th International Joint Conference on Natural Language Processing (EMNLP-IJCNLP)}, pp.\  3982--3992, 2019.

\bibitem[Ross(2023)]{ross2023introduction}
Sheldon~M Ross.
\newblock \emph{Introduction to Probability Models}.
\newblock Elsevier, 2023.

\bibitem[Sadasivan et~al.(2023)Sadasivan, Kumar, Balasubramanian, Wang, and Feizi]{sadasivan2023can}
Vinu~Sankar Sadasivan, Aounon Kumar, Sriram Balasubramanian, Wenxiao Wang, and Soheil Feizi.
\newblock Can ai-generated text be reliably detected?
\newblock \emph{arXiv preprint arXiv:2303.11156}, 2023.

\bibitem[Takezawa et~al.(2025)Takezawa, Sato, Bao, Niwa, and Yamada]{takezawa2025necessarysufficientwatermarklarge}
Yuki Takezawa, Ryoma Sato, Han Bao, Kenta Niwa, and Makoto Yamada.
\newblock Necessary and sufficient watermark for large language models, 2025.
\newblock URL \url{https://arxiv.org/abs/2310.00833}.

\bibitem[Yang et~al.(2023)Yang, Pan, Zhao, Chen, Petzold, Wang, and Cheng]{yang2023survey}
Xianjun Yang, Liangming Pan, Xuandong Zhao, Haifeng Chen, Linda Petzold, William~Yang Wang, and Wei Cheng.
\newblock A survey on detection of llms-generated content.
\newblock \emph{arXiv preprint arXiv:2310.15654}, 2023.

\bibitem[Zhang et~al.(2024)Zhang, Hussain, Neekhara, and Koushanfar]{zhang2024remarkllmrobustefficientwatermarking}
Ruisi Zhang, Shehzeen~Samarah Hussain, Paarth Neekhara, and Farinaz Koushanfar.
\newblock Remark-llm: A robust and efficient watermarking framework for generative large language models, 2024.
\newblock URL \url{https://arxiv.org/abs/2310.12362}.

\bibitem[Zhao et~al.(2023{\natexlab{a}})Zhao, Ananth, Li, and Wang]{zhao2023provable}
Xuandong Zhao, Prabhanjan Ananth, Lei Li, and Yu-Xiang Wang.
\newblock Provable robust watermarking for ai-generated text.
\newblock \emph{arXiv preprint arXiv:2306.17439}, 2023{\natexlab{a}}.

\bibitem[Zhao et~al.(2023{\natexlab{b}})Zhao, Ananth, Li, and Wang]{zhao2023provablerobustwatermarkingaigenerated}
Xuandong Zhao, Prabhanjan Ananth, Lei Li, and Yu-Xiang Wang.
\newblock Provable robust watermarking for ai-generated text, 2023{\natexlab{b}}.
\newblock URL \url{https://arxiv.org/abs/2306.17439}.

\bibitem[Zhu et~al.(2024)Zhu, Galjaard, Chen, and Chen]{zhu2024duwakdualwatermarkslarge}
Chaoyi Zhu, Jeroen Galjaard, Pin-Yu Chen, and Lydia~Y. Chen.
\newblock Duwak: Dual watermarks in large language models, 2024.
\newblock URL \url{https://arxiv.org/abs/2403.13000}.

\end{thebibliography}
